\documentclass[10pt,twocolumn]{article}
\usepackage{times}
\usepackage[utf8x]{inputenc}
\usepackage[final]{pdfpages}
\usepackage{t1enc}
\usepackage{graphicx}
\usepackage{textcomp}
\usepackage[T1]{fontenc}

\newcommand{\Section}[1]{\vspace{-8pt}\section{\hskip -1em.~~#1}\vspace{-3pt}}

\makeatletter
  \renewcommand{\@makecaption}[2]{\small #1: #2}   %%  #1= Zaehler, #2= Text
\makeatother

\begin{document}

\twocolumn[
%\centerline{\includegraphics[width=\columnwidth]{Copernicus_Meetings}}
\vskip10mm

\centering{\LARGE \bf Coherent dust cloud observed by three {\it Cassini} instruments}
\vskip10mm

% \begin{flushleft}
{\bf Emil Khalisi}\\
{Max-Planck-Institute for Nuclear Physics, Heidelberg, Germany}
%\end{flushleft}

%%%%%%%%%%%%%%%%%%%%%
%Please do not change/delete the following vskip command as it
%guarantees the correct distance between abstract information and abstract text.
\vskip10mm
]
%%%%%%%%%%%%%%%%%%%%%

\thispagestyle{empty}

\section*{Abstract}

We revisit the evidence for a "dust cloud" observed
by the {\it Cassini} spacecraft at Saturn in 2006.
The simultaneous data of 3 instruments are compared
to interpret the signatures of a coherent swarm of
dust that could have remained floating near the
equatorial plane.

\Section{Introduction}

Interplanetary dust clouds are local density enhancements
of particles of a specific mass type.
Such clouds might usually be relicts of a dissolved comet,
debris of an asteroidal collision,
ejecta from planets or moons,
and a few may also go back to jet streams from active
bodies,
or coronal mass ejections.
In the vast range of patterns, the characteristics of dust
will vary on all dimensions: size, density, mass,
lifetime, and more (see, e.g., Gr\"un {\it et al.} (2004)).
%\cite{gruen-2004}).

Kennedy {\it et al.} (2011) %\cite{kennedy-2011}
looked for a dust swarm in the
far-off field at Saturn ($>$100 Saturnian radii, R$_S$)
using {\it Spitzer} observations.
A large-scale cloud, that could be attributed to an
irregular satellite or other cosmic origin, was not
found definitely.
More recently, Khalisi {\it et al.} (2015) %\cite{khalisi-2015}
reported of one possible dust cloud detected as a persistent
feature in the impact rates of the Cosmic Dust Analyser
on the {\it Cassini} spacecraft.
We revisit their data in a broader context and complement
it with new evidence.
We synchronised the data by the Cosmic Dust Analyser (CDA),
the Radio/Plasma Wave Detector (RPWS),
and Magnetometer (MAG) to search for patterns that may
give evidence for a cloud of particles.
Unfortunately, the Plasma Spectrometer (CAPS) did not
provide sufficient output in the time slot considered.

\Section{Data Basis}

The key parameters of the instruments on the {\it Cassini}
spacecraft are stored in the MAPSview database.
We employed the following parameters for our study:
\begin{itemize}
\item CDA: impact rate $r^{\prime}_{\rm all}$ of all the
registered dust events per 64 seconds, see \cite{khalisi-2015};
%
%\item CAPS: density of ions $n_{\rm ion}$ (IMNT parameter);
%%
\item RPWS: qualitative radio signals in the frequency bands of
1 Hz, 10 Hz, 100 Hz, 1 kHz, and 10 kHz;
\item MAG: strength of the magnetic field |{\bf B}| plus its
three spacial components in the kronocentric solar-magnetospheric
(KSM) coordinate system.
%as well as spacecraft coordinates (MAG-SC);
%
%\item[+] TRAJ: additionally the current distance of {\it Cassini}
%from Saturn.
%
\end{itemize}
Most parameters are provided at 1-minute intervals of time,
except some very few cases when the instrument was out of
nominal operation.

The CDA data depend on the current instrument pointing
and reflect the density as well as other states of
the dust only partially.
RPWS and MAG are not reliant on the spacecraft attitude and
have the advantage of a continuous measurement of their
respective signals throughout the orbit.
In particular, the components of the {\bf B}-vector give
important clues to alignments of the magnetic field or
circular currents.

\Section{The dust cloud of DOY 203/2006}

\begin{figure}[ht]
\centerline{\includegraphics[width=\columnwidth]{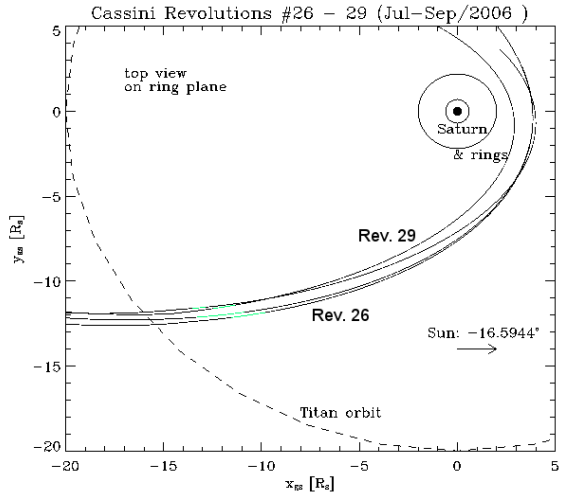}}
\caption{Trajectory of {\it Cassini} from outer regions
heading to its perikronium. The region of the dust cloud
is highlighted green.}
\label{fig:orbit-overlap}
\end{figure}

The {\it Cassini} spacecraft just changed its sequence
of orbits from equatorial to inclined trajectories
after a targeted flyby at Titan (T16) on DOY 203.02.
It set in for four consecutive revolutions (\#26--29,
Fig.~\ref{fig:orbit-overlap})
out of the equatorial plane to traverse almost the same
spot in the Saturnian space.
The inclination to the ring plane was about 15$^{\circ}$
for these revolutions (not seen in that projection).
The ring plane of Saturn was crossed at DOY 203.11
in a distance of 20.126 R$_S$.
The green segments mark the region of the supposed
dust cloud.

%%%
%%% Solar latitude for Revolutions #....
%%% ------------------------------------
%%% Rev 25 (DOY 179):
%%% Rev 26 (DOY 203): -16.59439
%%% Rev 27 (DOY 227): -16.27620
%%% Rev 28 (DOY 251): -15.94156
%%% Rev 29 (DOY 266): -15.75277
%%%

\begin{figure}[ht]
\centerline{\includegraphics[width=\columnwidth]{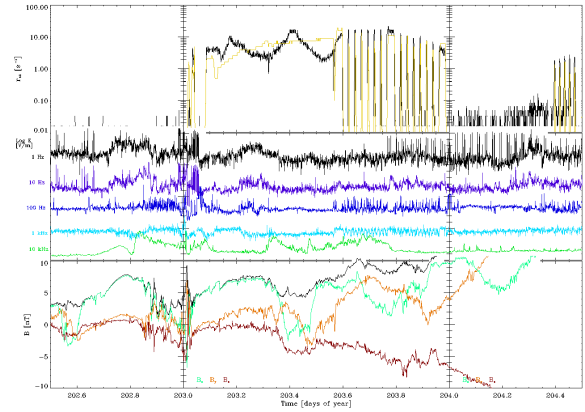}}
\caption{Data comparison of three Cassini
instruments at Revolution \#26 in July 2006.
{\it Uppermost panel:} Dust impact rate $r_{\rm all}$
(black line) and the sensitive area (yellow) of the CDA
exhibited to the Kepler-RAM.
%{\it Second panel:} Ion density $n_{\rm ion}$ of the
%CAPS instrument.
{\it Middle panel:} Five frequencies of the radio
and plasma data (RPWS).
{\it Bottom panel:} Magnetic field strength |{\bf B}|
(black) as well as the 3 B-components in the frame of
kronocentric-solar-magnetospheric coordinates.}
\label{fig:cloud2006203}
\end{figure}

Figure~\ref{fig:cloud2006203}
shows the time-depended signals of the three
instrumental parameters in July 2006.
The peak in the dust rate at DOY 203.40 coincides with
a depression of |{\bf B}| (black line),
and, in particular, the B$_x$-component (blue) changes
its direction.
The B$_y$-component (orange) exhibits a decline from
positive to negative values and rising back again
showing a clear sign of rotation as a ring current
was hit.
At time $T$ = 203.50, the previous {\bf B}-values were
restored acting as if separated by a bulkhead.
Goertz (1983) reported of similar plasma tubes
from both {\it Voyager} flybys.
He interpreted that as ``plasma blobs'' breaking off
the magnetospheric sheet.

The RPWS data support the thought of a ``magnetic bubble'',
for it shows a remarkable tranquility in the
100 Hz and 1000 Hz band
(third and fourth curve of RPWS in the middle panel).
A dozen "negative peaks" appear in the 1 Hz band
(top black curve).
These peaks can be considered as micron-sized dust
grains impacting on the spacecraft.
Similar features in the 10-kHz-band were discussed
by Kurth {\it et al.} (2006). % \cite{kurth-2006},
%that happen on the scale of milliseconds of time.
Moreover, the CDA registered twice as much impacts
on the instrument housing than on its sensitive areas,
meaning that many particles did not enter from
the Kepler-RAM direction.
Time stamps of the most conspicuous features are
given in Table~\ref{tab:maincloud}.

\begin{figure}[ht]
\centerline{\includegraphics[width=\columnwidth]{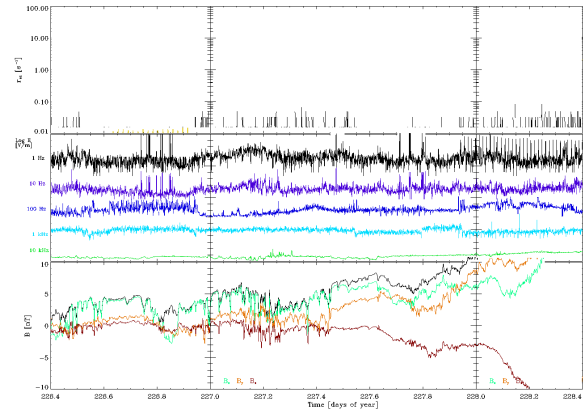}}
\caption{Data of CDA, RPWS, and MAG for
Revolution \#27.
The main feature of the supposed dust cloud
resides from T = 227.11 to 227.42.}
\label{fig:cloud2006227}
\end{figure}

\begin{figure}[h]
\centerline{\includegraphics[width=\columnwidth]{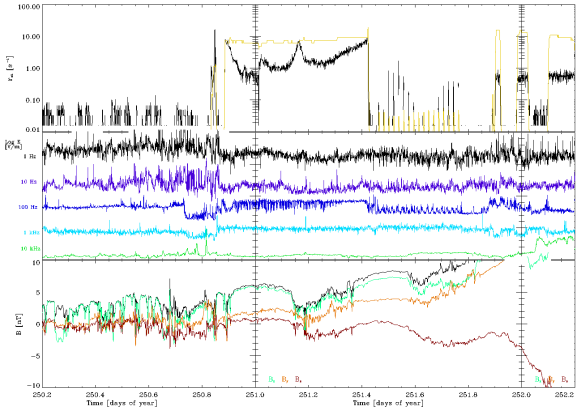}}
\caption{Panels same as in Figures~\ref{fig:cloud2006203}
and \ref{fig:cloud2006227} for Revolution \#28.
The ring plane crossing happened on 250.89.
}
\label{fig:cloud2006251}
\end{figure}

The same pattern repeated at the next two passages when
the spacecraft passed almost the same spot of space.
During the Revolution \#27 (Fig.~\ref{fig:cloud2006227}),
the CDA pointed to an unsuitable direction, but the
countersink of the magnetic field remained.
%The 10-kHz-band of RPWS also suggests some kind of
%anomaly from 227.00 to 227.31 with higher signal
%intensity.
%The data of this revolution shows large uncertainties
%though.
---
At the third return the signals resemble DOY 203
again (Fig.~\ref{fig:cloud2006251}).
In spite of some changes among minor features,
important characteristics of the first passage can
be identified.
From the time stamps of entering and leaving, the
radial extent of that ``cloud'' or ``blob'' can be
estimated to $\approx$82,500 km or 1.36 $R_{\rm S}$.
Its $z$-location is found 1 $R_{\rm S}$ below the
ring plane.
---
At the fourth passage on DOY 266/2006 (not shown),
the CDA data displays even stronger deviations,
and the allocation of that cloud turns out uncertain.

%\onecolumn
%
%
%\vskip10mm
%\nopagebreak

%%%%%%%%%%%%%%%%%%%%

%\clearpage

%%%%%%%%%%%%%%%%%%%%

The MAG-data suggests that the cloud has broadened
by 15,000 km from Revolution \#26 through \#28.
It could also have moved outwards from 18.5 $R_{\rm S}$
to 18.9 $R_{\rm S}$, though this might be an effect
of the spacecraft hitting the cloud at different
parts.
The slightly  different trajectory can also be the
reason for the discrepancies in the data at the
second return (DOY 227).
The increase of size, however, is supported by the drop
of |{\bf B}| by 2--5 nT.
The magnetically ``quiet'' region in front of the
onset of the cloud ($\approx$250.92--251.14) has also
diminished by 1 nT.
The bulk of solid particles (CDA) appears compressed
and shifted further from the deepest point of the
magnetic field.
This deviation seems to grow.

There exist indications for two more clouds during the
Revolution \#26:
from T = 202.95 to 203.05 and from T = 203.65 to 203.95.
Both can be re-discovered at the next but one flyby:
in Revolution \#28, they peak around 250.80 and 251.50,
respectively.

%%%%%%%%%%%%%%%%%%%%

\begin{table}[ht]
\caption{Time stamps $T$ of the dust cloud as seen by different
instruments.
$D$ = distance of the pattern peak from Saturn.
Brackets reflect very uncertain data.}
\label{tab:maincloud}
\vskip4mm
\centering
\begin{tabular}{lccc}
\hline
Instr. & $T$ (enter) & $T$ (leave)& $D$ [$R_{\rm S}$] \\
\hline
\multicolumn{3}{c}{Revolution 26}\\
CDA              & 203.32 & 203.52  & 17.69\\
MAG              & 203.33 & 203.50  & 17.64\\
RPWS$_{\rm 1 Hz}$  & 203.36 & (203.53)& \\
RPWS$_{\rm 100 Hz}$& 203.36 & 203.56  & 17.64\\
RPWS$_{\rm 10 kHz}$& 203.31 & 203.56  & 17.71\\
%RPWS$_{\rm 100 kHz}$&203.32 & (203.53)\\
\hline
\multicolumn{3}{c}{Revolution 27}\\
CDA              &(227.10)&(227.39) & (18.78)\\
MAG              & 227.11 & 227.42  & 18.93\\
RPWS$_{\rm 1 Hz}$  &(227.06)& (227.28)&\\
RPWS$_{\rm 100 Hz}$& 227.09 & 227.39  & 18.70 \\
RPWS$_{\rm 10 kHz}$& (227.00)& 227.31 & \\
%RPWS$_{\rm 100 kHz}$&227.10 & 227.39\\
\hline
\multicolumn{3}{c}{Revolution 28}\\
CDA              & 251.11 & 251.25  & 18.48\\
MAG              & 251.13 & 251.29  & 18.34\\
RPWS$_{\rm 1 Hz}$  &(251.13)&(251.26)& \\
RPWS$_{\rm 100 Hz}$& ---    & ---    & \\
RPWS$_{\rm 10 kHz}$& ---    & ---    &\\
%RPWS$_{\rm 100 kHz}$&(251.09)& 251.28\\
\hline
\end{tabular}
\end{table}

\Section{Discussion}

It is still not certain whether or not such coherent
swarms of dust can exist for long in the magnetosphere
of Saturn.
From the theoretical point of view, small dust particles
are quickly ionised (UV-radiation, solar wind) and
carried away by the magnetic field.
Various other effects like shock waves, gravitational
drags, evaporation, and Kepler shear will also lead to
a disruption of the cloud.
It will preferably be the debris larger than
$\approx$1 mm in size that may endure several crossing
times, $r/v$,
where $r$ is the projected radial distance and $v$ the
velocity dispersion of the particles.
Such larger particles are not recorded though.
However, the most intriguing evidence for the supposed
clouds of small dust particles comes from the CDA.
Further detailed investigation is envisaged and more
cases like this one have to be found.
%before the existence
%of such dust clouds can be confirmed.
Since the trajectory of {\it Cassini} changes too
frequently, it will be challenging to trace a
particular cloud for more than two or three revolutions
of the spacecraft.
Within that period of time, we expect the cloud to change
its shape.
More evidence for the dust clouds is in preparation.

\section*{Acknowlegements}

We thank Ralf Srama and Georg Moragas-Klostermeyer for
a helpful discussion.

\end{document}